\documentclass[aps,pre,longbibliography,notitlepage,floatfix,nofootinbib,superscriptaddress]{revtex4-1}

\usepackage{bm,amsmath,amssymb,graphicx,stmaryrd,float,subcaption,upgreek,sidecap,dsfont}
\usepackage[abs]{overpic}
\usepackage{verbatim}
\usepackage{comment,footnote}
\usepackage[format=plain,labelfont={bf,small},textfont=small,justification=raggedright,singlelinecheck=false]{caption}

\usepackage[export]{adjustbox}

\graphicspath{{./Figures_paper/}}

\usepackage[colorlinks=true]{hyperref}
\hypersetup{
	colorlinks = true,
   linkcolor=blue,
   urlcolor=blue,
   citecolor=red
}

\begin{document}
	\title{Impact-induced acceleration by obstacles}
  \affiliation{Department of Biomedical Engineering and Mechanics,}
	\affiliation{Department of Physics,}
	\affiliation{Center for Soft Matter and Biological Physics, \\ Virginia Polytechnic Institute and State University, Blacksburg, VA 24061, U.S.A.}
	\author{N. A. Corbin}
	\affiliation{Department of Biomedical Engineering and Mechanics,}
	\author{J. A. Hanna}
	\email{hannaj@vt.edu}
	\affiliation{Department of Biomedical Engineering and Mechanics,}
        \affiliation{Department of Physics,}
        \affiliation{Center for Soft Matter and Biological Physics, \\ Virginia Polytechnic Institute and State University, Blacksburg, VA 24061, U.S.A.}
	\author{W. R. Royston}
	\affiliation{Department of Biomedical Engineering and Mechanics,}
	\author{H. Singh}
	\affiliation{Department of Biomedical Engineering and Mechanics,}
	\author{R. B. Warner}
	\affiliation{Department of Biomedical Engineering and Mechanics,}

\date{\today}

\begin{abstract}
We explore a surprising phenomenon in which an obstruction accelerates, rather than decelerates, a moving flexible object.
It has been claimed that the right kind of discrete chain falling onto a table falls \emph{faster} than a free-falling body.
We confirm and quantify this effect, reveal its complicated dependence on angle of incidence, and identify multiple operative mechanisms.
Prior theories for direct impact onto flat surfaces, which involve a single constitutive parameter,
match our data well if we account for a characteristic delay length that must impinge before the onset of excess acceleration.
Our measurements provide a robust determination of this parameter.
   This supports the possibility of modeling such discrete structures as continuous bodies with a complicated constitutive law of impact that includes angle of incidence as an input.
\end{abstract}
\maketitle

\vspace{-0.75cm}
\section{Introduction}

The impact and interaction of flexible discrete and continuous structures with obstacles is an area full of  counterintuitive physics, from the classic textbook example of a falling ladder \cite{FreemanPalffy-Muhoray85}, to unwrapping \cite{Cambou12, Brun16} and the buzz of musical instruments \cite{Raman21, Burridge1982}, to chatter in mechanical systems \cite{BuddDux94, DankowiczFotsch17} and the death clatter of expensive rectangles \cite{Goyal98-1, Goyal98-2, iphonedroptest}, 
to less well understood phenomena such as complicated velocity-restitution relationships \cite{StoianoviciHurmuzlu96, King11, Muller13}, deformable water-walking projectiles \cite{Belden16}, and the chain fountain \cite{MouldSiphon2, MillerYouTube, JuddVimeo, Biggins14, Virga14}, 
 a subject of much recent amateur experimentation.
In this note, we confirm and quantitatively explore the surprising appearance of an excess acceleration induced when an extended, flexible object falls onto a rigid surface.  The effect depends on the discreteness of the object at small scales, and is thus an example of ``microscopic'' physics manifesting in the large.

The mechanics have obvious connections to cable laying \cite{Zajac57} and impact oscillations of mooring lines \cite{Gobat2001}.  We also anticipate relevance to the phenomenologically rich subject of liquid rope coiling \cite{Ribe12, BarnesWoodcock58, Morris08, BlountLister11, Brun15} and related processes. Even small forces induced at a nozzle by the substrate during spinning of polymer fibers may affect the stability of the resulting jet \cite{PetrieDenn76, Renardy06, CruickshankMunson81, Tchavdarov93}.
Studying the dynamic contact behavior of extended objects may help us manipulate flexible robotic casters \cite{Hill15} and understand passive reconfiguration and obstacle avoidance in organismal biology, as when cockroaches navigate tall compliant grass \cite{Li15}, or snakes land on tree branches \cite{Socha11}.  

\begin{figure}[h!]
	\begin{minipage}{3.75cm}
		\includegraphics[width=\linewidth,valign=t]{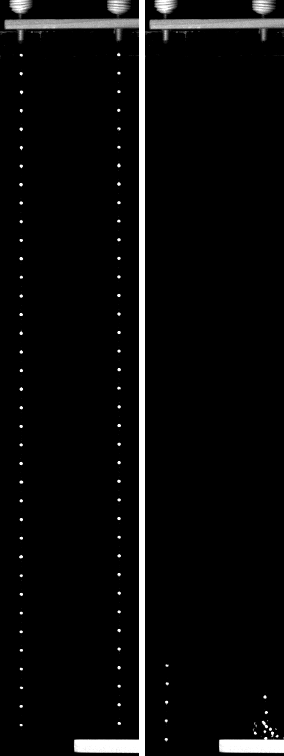}
		\label{Fig:distant_shot}
	\end{minipage}
		\hspace{0.01cm}
	\begin{minipage}{10cm}
		\includegraphics[width=0.495\linewidth,valign=t]{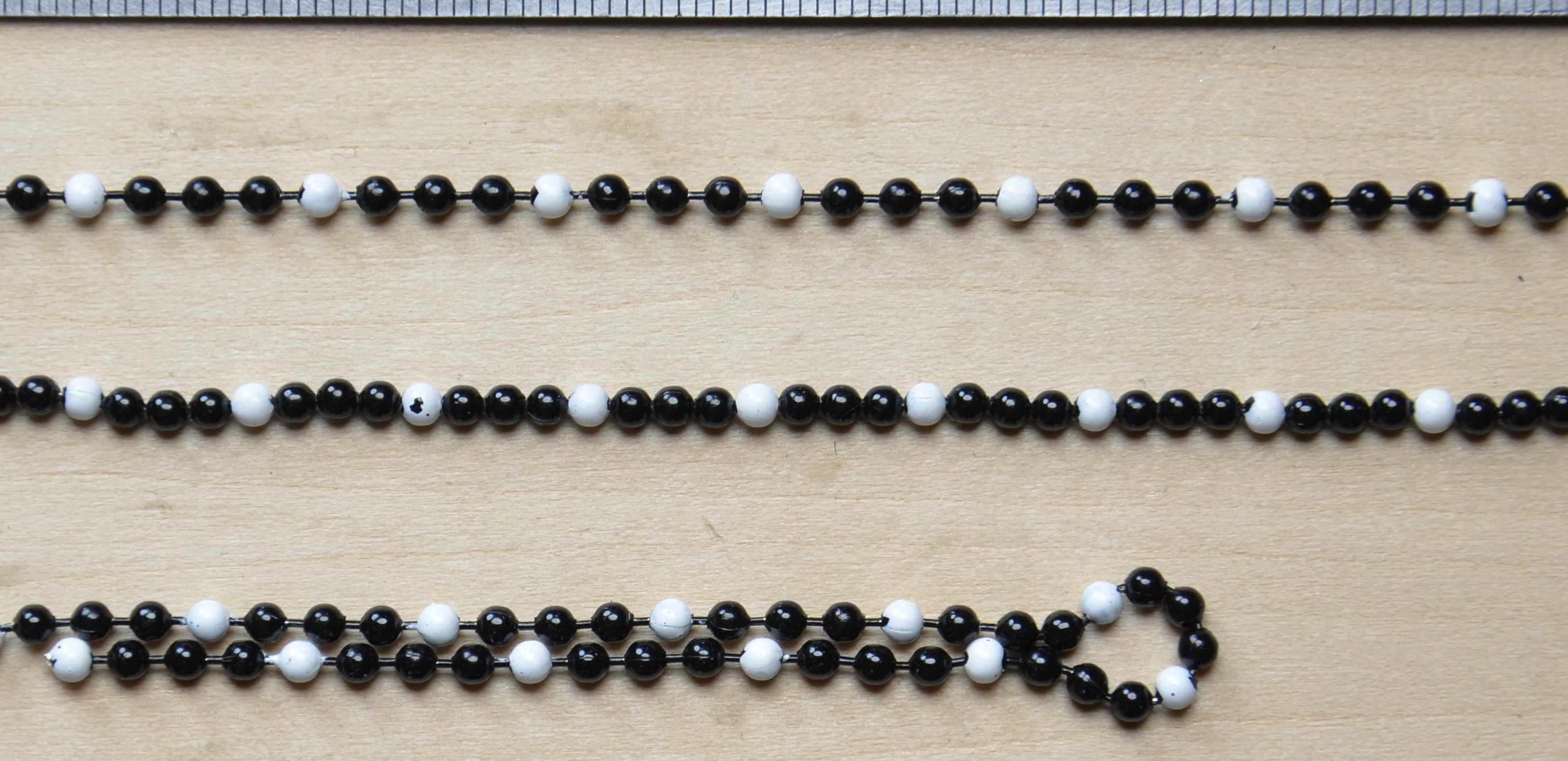}\label{Fig:closeup_chain}
		\hspace{0.01cm}
		\includegraphics[width=0.47\linewidth,valign=t]{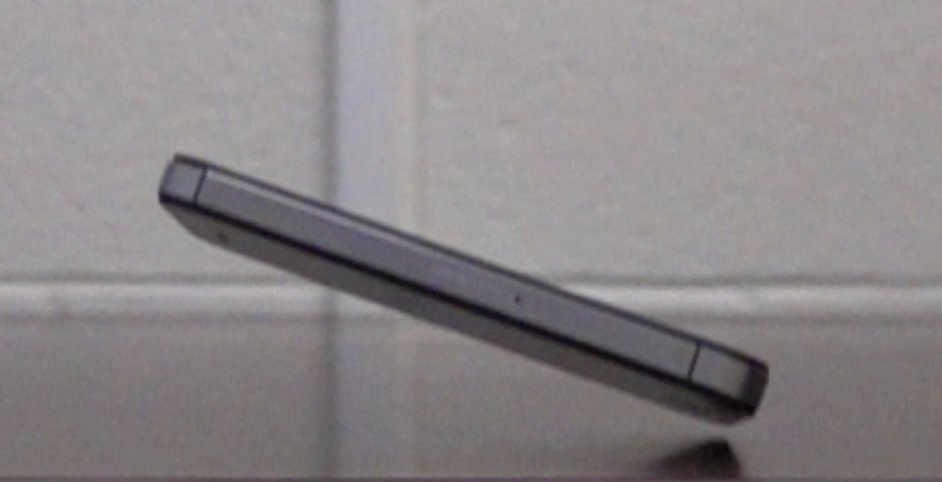}\\
		\vspace{0.75cm}
		\includegraphics[width=.96\linewidth,valign=b]{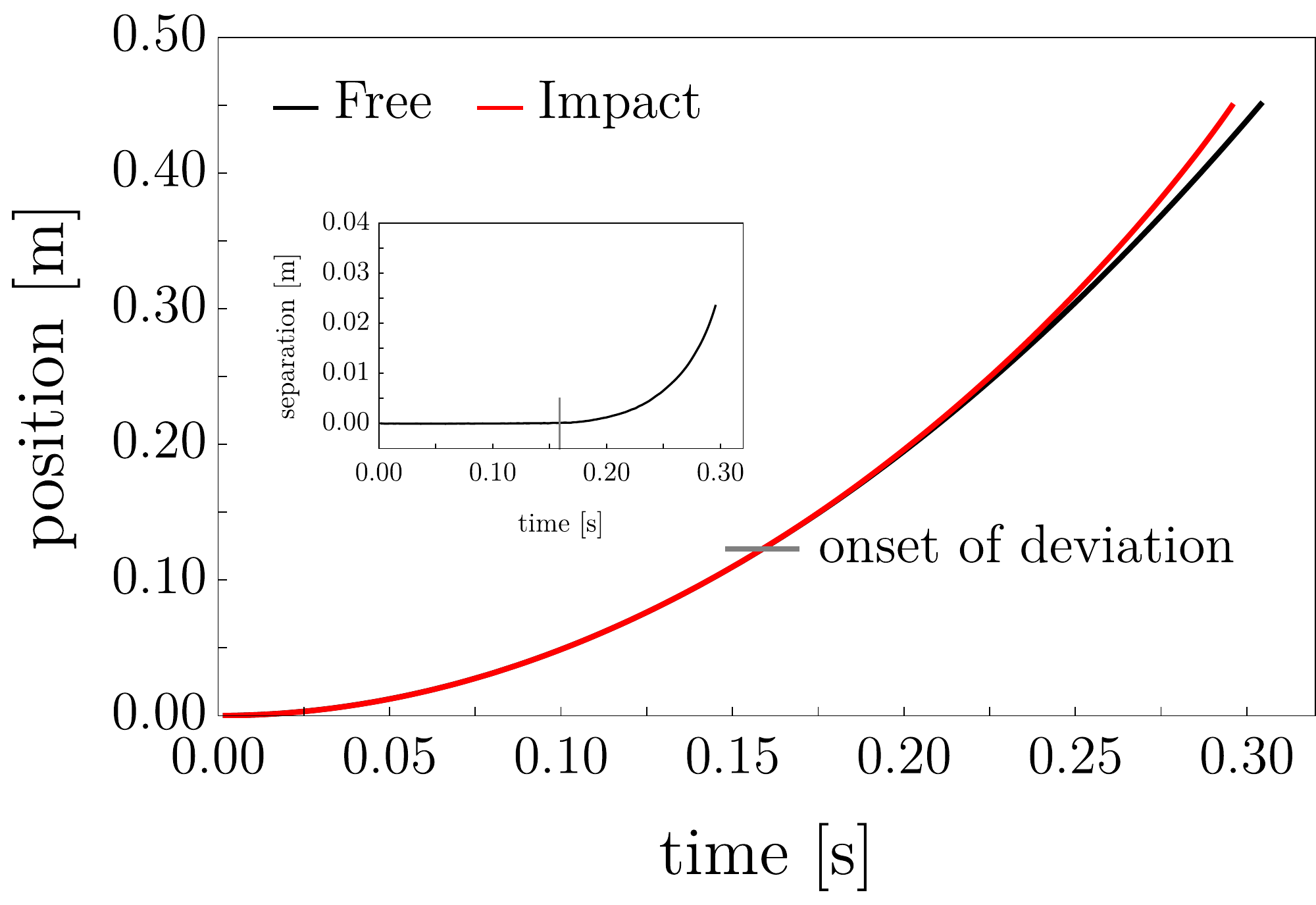}\label{Fig:position_curves}
		\vspace{0.25cm}
	\end{minipage}
	\caption{Clockwise from left:  Snapshots before and after dropping two identical ball chains (black with white markers on every fourth ball), showing that an impeded chain falls faster than a free-falling chain ; Closeup of chains in extended, compressed, and bent states ; When the lower right edge of a falling rectangle (an iPhone 4S) hits a table, the upper left edge will accelerate downward ; Positions of markers near the top ends of freely falling (black) and impeded (red) chains, and their difference (inset).  The onset of deviation is marked as the time when the separation exceeds the threshold value of 0.1\,mm.
}
	\label{Fig:Experimental_setup}
\end{figure}

Our experiments were performed to resolve questions raised by a pair of recent papers.  In the course of a related study of impact forces, Hamm and G{\'{e}}minard \cite{HammGeminard10} observed what appeared to be an anomalous extra downward acceleration induced in a ball chain (also known as a bead chain) falling into a pile, as compared with a freely falling steel ball.  
Ruina and co-workers \cite{Grewal11} objected that to avoid differential drag effects, experiments should be performed with two identical flexible bodies, one falling free and one impeded by an obstacle.
However, they then proceeded to perform this experiment not with a ball chain, but with two other types of chain.  One was a link chain (also known as a cable chain) consisting of linked toroidal loops that become mechanically disconnected when subjected to a compressive force--- this chain experienced no extra acceleration.  The other was a specially-designed ladder-like chain consisting of slanted rods, which did experience an extra acceleration.
The mechanism demonstrated in the latter case is simple to understand--- when one end of a falling rigid link impacts a surface at some angle, the other end accelerates towards the surface because of the induced rotation, as is about to happen to the rectangular phone in Figure \ref{Fig:Experimental_setup}.  From a macroscopic point of view of a chain composed of such small links, this looks like a force ``pulling'' downward; the reverse mechanism has been proposed as the source of the pickup ``push'' of a chain fountain \cite{Biggins14}.
However, the ball chain itself has link-like elements, and a somewhat larger semi-rigid length scale associated with a minimum bending radius that influences a coiling radius during steady-state impingement, as pointed out by Hamm and G{\'{e}}minard.
A conclusive apples-to-apples experiment involving two ball chains remained unperformed until the present study \cite{Corbin17}.  Another unresolved issue in these papers is the role of drop height, and thus velocity, of the impacting chains.

In the present work, we confirm the existence of the effect in a ball chain and obtain quantitative data on trajectories and constitutive behavior, revealing multiple, at times competing, mechanisms and a complicated dependence on the angle of the impacted surface.
We resolve some outstanding questions, and lend support to conjectured explanations and theoretical modeling from the work of both aforementioned groups as well as a dissipative shock model proposed by Virga \cite{Virga15}.


\section{Experimental details}

We use ball chains (\#3 black epoxy coated steel ball chain, Ball Chain Mfg.\ Co., Inc., Mount Vernon, NY) with a nominal and measured ($\pm\,$0.05\,mm) ball (bead) diameter of 3/32\,in (2.4\,mm). 
Every fourth bead is turned into a white Lagrangian marker using acrylic high gloss 
touch up paint, after which the chain weight is approximately 0.029\,g per bead and link unit. 
The chain is compressible, with an extended center-to-center inter-bead distance of 3.36$\,\pm\,$ 0.05\,mm under tension and a compressed distance of 2.46$\,\pm\,$0.05\,mm, and has a minimum radius of approximately 3\,mm, involving 8 or 9 beads, with a maximum bending angle of 47$\,\pm\,$2$^\circ$ between adjacent beads (Figure \ref{Fig:Experimental_setup}).
There are small dumbbell-shaped connectors that can slide into the balls to allow compression, but will not allow extension beyond the extended distance.
Bending and compression are somewhat coupled, as can be seen in the figure.


For a typical test, two identical chains are dropped simultaneously, one free and one impeded by an obstacle, as shown in Figure \ref{Fig:Experimental_setup}.
For several data sets taken under identical conditions, only the data from one free chain was used for comparison with impacting chains.  Obstacles were either a flat (horizontal) wooden surface or sloped surfaces made from printed PLA (Ultimaker 2 Extended+, Geldermalsen, the Netherlands) using a 0.1 mm layer thickness.  Based on our own and others' \cite{HammGeminard10} observations, we assume that the choice of surface material does not affect the results.
Before release, the chains hang vertically from electromagnets for 20 minutes in a room with ventilation fans shut off, 
 to damp out vibrations.  We use as weak a current as possible to avoid residual magnetic effects when power is cut to release the chains.
We record trajectories of beads and determine the vertical separation between beads of the same initial height on free and impeded chains.  The data of most interest is from near the upper end of the chains.  To avoid complications from end orientation effects due to interaction between the shape of the chain elements and the electromagnetic nail tips on which they are held (bent ends are visible in some videos), we use data from the second to last white marker, six beads from the end.
Velocities are essentially purely vertical and planar until beads approach very close to the impact point, at which time the data is truncated.  There is a potential one-sided error of about 2.9\,mm for final separations arising from our choice of where to truncate our data for each data set; we indicate this on the relevant figures independently of other sources of error contributing to the plotted error bars.  
The falls are recorded using a FASTCAM Mini UX100 camera (Photron, San Diego, CA), 
with an AF Nikkor 50\,mm f/1.4 D lens (Nikon, Tokyo, Japan), at a rate of 3200 frames per second.
Particle tracking of white markers is achieved using the PTVlab tool \cite{PTVlab}, along with some additional processing to allow for preservation of individual marker labels despite entrance or exit of markers from the field of view.
For much of the analysis, we use data resampled using every third point in time, and subsequently smoothed using two iterations of a three-point moving average. 
PTVlab evaluates the centroids of particles using a Gaussian mask \cite{PTVlab}. At the largest recording distances we used, a typical particle comprises about 15 pixels, and the pixel size is 0.72-0.73\,mm, as determined by using known chain lengths or other fiduciary markings.  This results in a vertical position resolution, as we could clearly observe in the aliasing of obtained data, of about 0.1\,mm.  We use this as a minimum threshold for interpreting data, noting that it is much less than half a pixel. 
A significant source of error arises because of uncertainty in synchronizing the drop times of the two chains, which is found to be on the order of the unresampled frame rate value of 1/3200\,s. 
Errors in time zeroing lead to errors in separation curves, obtained by subtracting the impeded curves from the free ones, and appear as an initial linear trend superimposed on the real nonlinear separations.
For some trials, we expect some kind of nonlinear separation to occur from the very beginning.  We estimate error in these tests to be less than half a millimeter error over a typical drop time, based on the worst case linear separation errors due to mistiming.
This is about 5\% of typical measured separations, and considerably less than the one-sided truncation error.
However, for many trials, there is a delay in the onset of nonlinear separation, so the initial drop behavior during which there is no separation can be used to align the two trajectories; any initial linear portion of the separation data is subtracted from the entire curve.
This improves errors in separation, but the main purpose of this process is to improve the estimates of when initial separation occurs, a much more uncertain measurement.
The error here is based on how much of the initial trajectory is used to re-zero the separation curve.  We choose what appears to be a reasonable interval, but also recalculate the data over a range of intervals and use the maximum and minimum as error estimates. 
Often there is a clean plateau of consistent results, but some cases show greater variability.  The process is automated except for a single point, the upper outlier of the right plot in Figure \ref{Fig:Delay}, for which error was estimated by inspection of the data.
Because of the rezeroing process and associated errors, we cannot rule out the possibility of a slight slowing down of some of the impacting chains around the time of impact.

As a calibration check, we used least-squares fits to parabolas of data from about 40 free-falling chains to estimate the gravitational acceleration in the lab, obtaining a mean and standard deviation of 9.78$\,\pm\,$0.01\,m/s$^2$.  This implies a consistent underestimation bias, equivalent to about one pixel over the relevant chain lengths of approximately 700 pixels, as compared with 
either estimates using the surface gravity prediction tool from the National Geodetic Survey \cite{NGSgravity} (9.797\,m/s$^2$) or precise measurements using a Kater's pendulum \cite{Schweizer1952} in a nearby building 75 years ago (9.7976$\,\pm\,$\,0.00128 m/s$^2$). 
Errors in vertical alignment, even of several pixels, would not be enough to create such a bias.
We note that the error in measured separations between simultaneous drops or drops performed under the same filming conditions is likely to be less than that implied by this bias.

\section{Results} 

Our first set of experiments confirms the existence of an extra acceleration for a ball chain dropped onto a flat surface and coiling into a pile.  
Coils on flat surfaces 
involve around 11 beads, so are only slightly larger than the minimum bending radius.
As seen in Figure \ref{Fig:Experimental_setup} and the video \texttt{0deg.mp4} \cite{videos}, the effect manifests as a final end separation of a bit more than 2\,cm for a simultaneously dropped pair of half-meter long chains.  A slight difference in release times for the two chains can also be detected in the initial stages of this video. 

\begin{figure}[h!]
	\begin{minipage}{8.4cm}
		\includegraphics[width=0.96\textwidth]{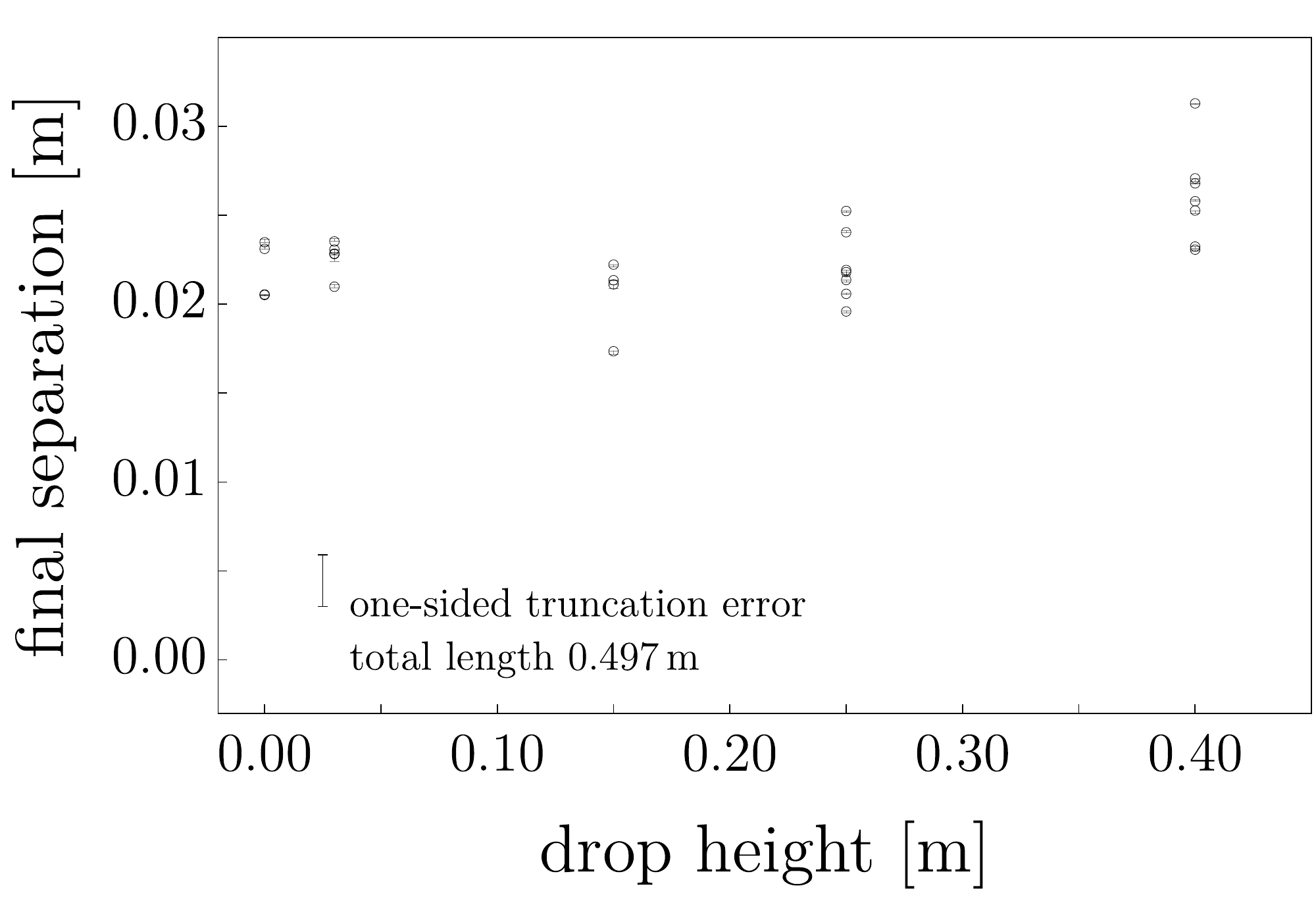}
	\end{minipage}
	\begin{minipage}{8.4cm}
		\includegraphics[width=0.96\textwidth]{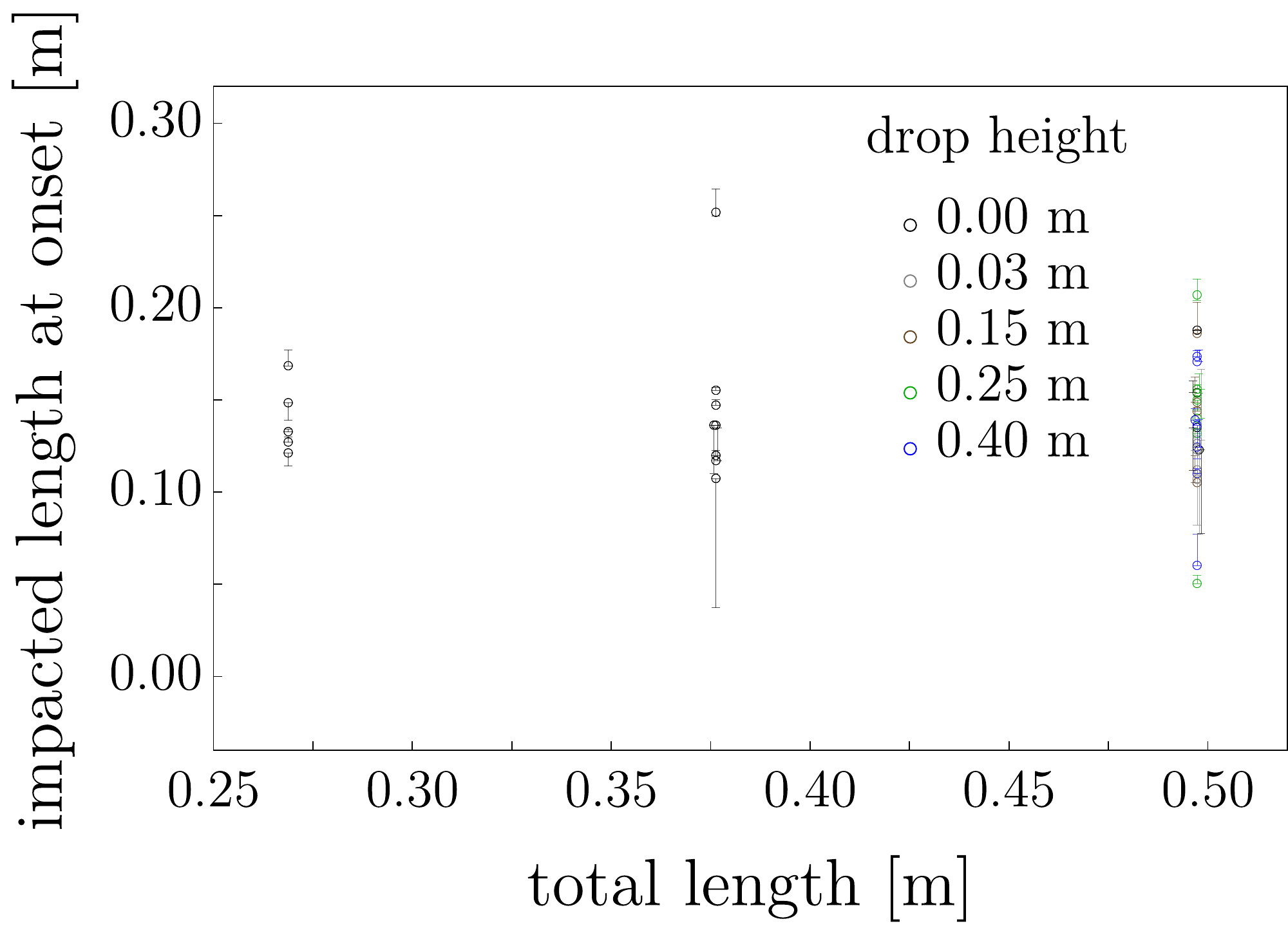}
	\end{minipage}	
\caption{Left: Final separation between beads near the ends of free falling and impacting long chains dropped from different heights.  Right: Deviation from free fall begins when a characteristic length of chain has already impacted, across chains of different lengths dropped from different heights.  Some data points and error bars are given slight lateral shifts to avoid overlap. Drop height errors are $\,\pm\,$4\,mm and total length errors are sub-millimeter.}
\label{Fig:Delay}
\end{figure}

\begin{figure}
	\begin{minipage}{12cm}
		\includegraphics[width=\textwidth]{{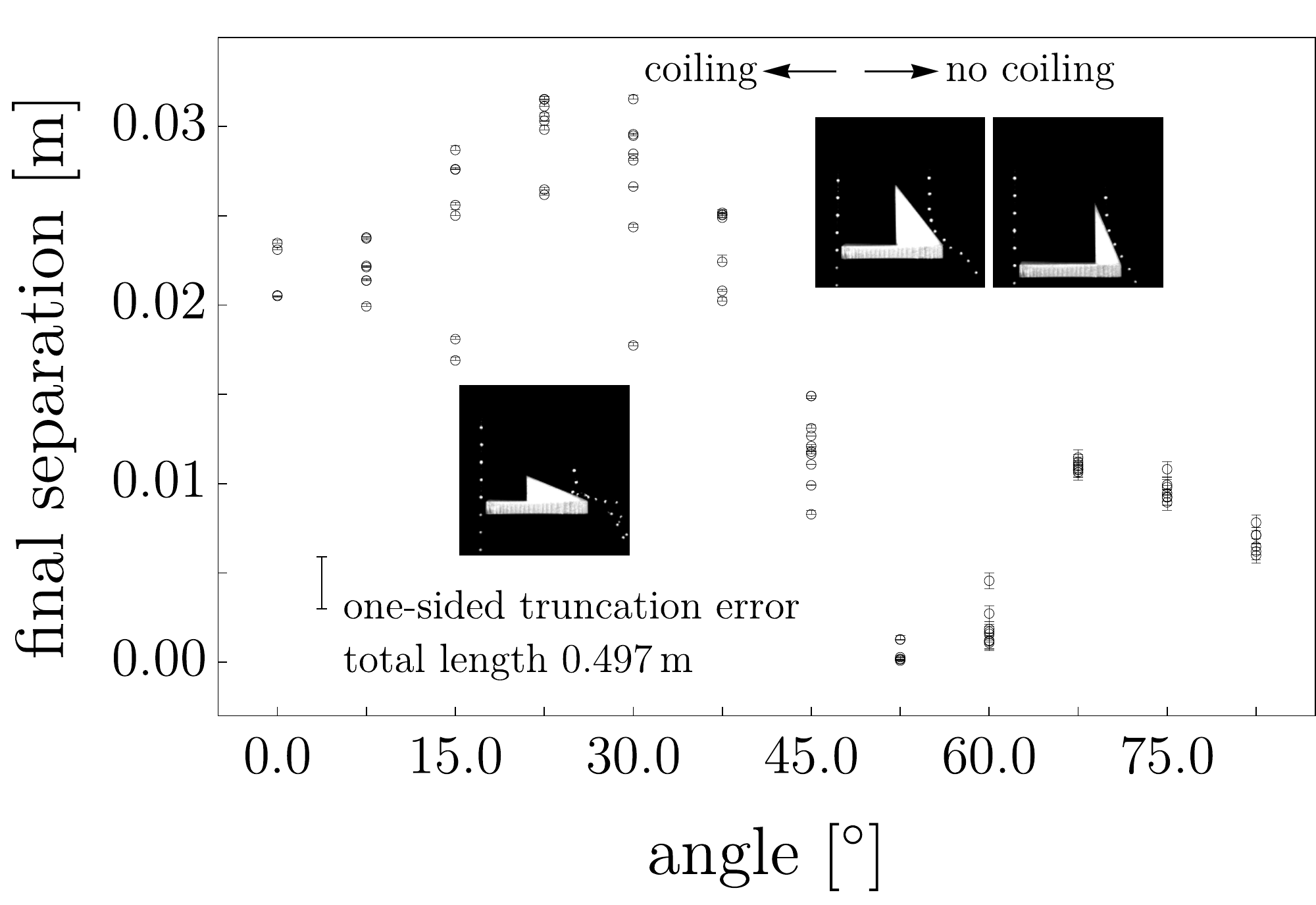}}
	\end{minipage}
	\caption{Final separation between beads near the ends of free falling and impacting long chains dropped onto surfaces of different inclinations, with inset images at $22.5^\circ$ (maximum effect, coiling), $52.5^\circ$ (no effect, no coiling), and $67.5^\circ$ (small effect, no coiling).}
	\label{Fig:Separation_vs_angle}
\end{figure}

The impression given by the high speed films is that of a delayed onset of deviation from free-fall, with the extra acceleration kicking in some time after impact.  This is consistent with the recorded data, of which Figure \ref{Fig:Experimental_setup} shows an example.  There are several possible reasons why one might expect a transient such as this.  The effect could depend on a threshold in velocity or in proximity to the free end of the chain (affecting the tension or its gradient), a time scale for propagation of transverse waves, a time or distance scale for the initiation of coiling, or a distance scale set by the necessity for steric interactions with a pre-formed pile of chain.
To eliminate some of these possibilities, we performed drops of chains of end bead center-to-center lengths (under tension) of $L$\,=\,497$\,\pm\,$0.6, 376$\,\pm\,$0.4, and 269$\,\pm\,$0.3\,mm (149, 113, and 81 beads) from heights of 0, 30, 150, 250, or 400$\,\pm\,$4\,mm from the bottom of the chain to the impacting surface.
An example of a long chain dropped from a height can be seen in the video \texttt{0degheight.mp4} \cite{videos}, and a summary of relevant results can be seen in Figure \ref{Fig:Delay}.
We observed no clear connection between the onset of deviation 
and velocity or distance from the free end at impact, and no characteristic delay time. 
There may be a weak correlation between drop height (impact velocity) and total separation, but if so, this does not arise from an earlier onset of deviation, but rather from a dependence of incremental rate of deviation on velocity, something predicted by existing theories \cite{HammGeminard10, Grewal11, Virga15}. 
However, we find a characteristic delay \emph{length}, such that a length of approximately 0.15\,m of chain (45 beads) experiences impact before the onset of deviation. 
Our initial reaction was that this seems consistent with the need for the formation of a pile of a given size to provide steric interactions that lead to the observed extra accelerations.  However, this explanation is incorrect.  Attempts to organize the pile formation in containers, or the planting of a pre-formed pile of chain under the impinging chain, did not lead to any noticeable changes in behavior.  Use of a sloping impact surface, such that impact coils are swept out of the way of newly impinging chain, also does not eliminate, and can even enhance, the extra acceleration after a similar initial delay.

In fact, the deviation from free-fall displays a complicated bimodal dependence on the slope of the impact surface, as shown in Figure \ref{Fig:Separation_vs_angle}.  
Half-meter length chains were dropped onto ramps with angles of inclination ranging from $0$-$82.5^\circ$ at increments of $7.5^\circ$.
Examples can be seen in the videos \texttt{22p5deg.mp4} , \texttt{52p5deg.mp4} , and \texttt{67p5deg.mp4} \cite{videos}.
At low angles, we observed an increase, and subsequent decrease, of the effect with increasing impact angles. 
Somewhere between 45$^\circ$ and 52.5$^\circ$, the chains no longer coil; it's not clear whether there is any connection between this transition and the maximum bending capacity of the chain, which would correspond to a tilt angle of $90 - 47 = 43^\circ$.
At 52.5$^\circ$, the effect disappears entirely, only to return at higher angles, where the mechanism of impact is simpler than coiling, involving a straightforward falling link mechanism.  Here, the resulting end separation due to extra acceleration is significantly less than that at lower angles.
Attempts to change the frictional properties of the ramps or place impediments at the ends of the ramps to induce pile-ups did not lead to any noticeable changes in behavior.

\begin{figure*}[t]
	\begin{minipage}{16.8cm}
		\includegraphics[width=0.48\linewidth,valign=t]{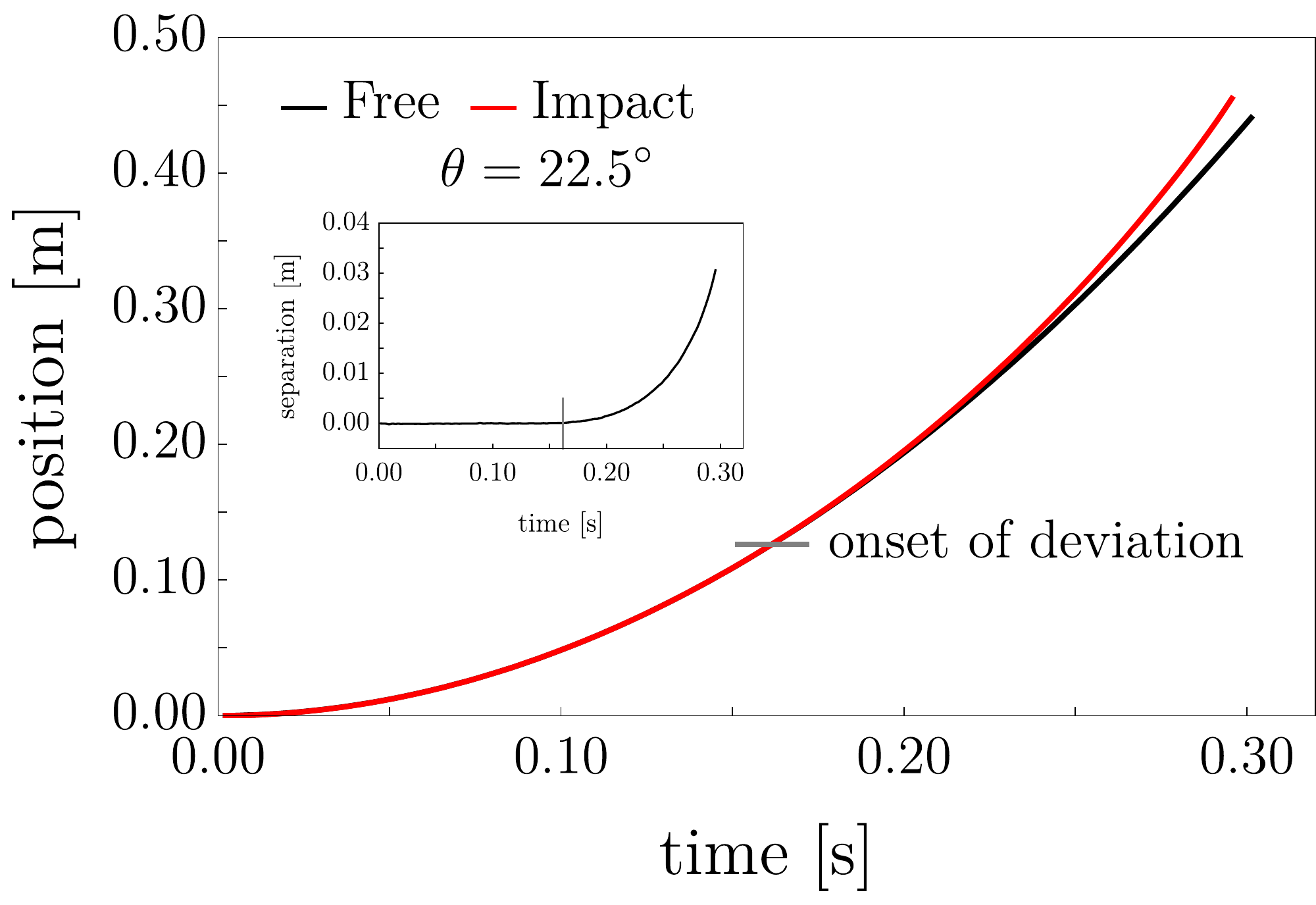}
		\hfill
		\includegraphics[width=0.48\linewidth,valign=t]{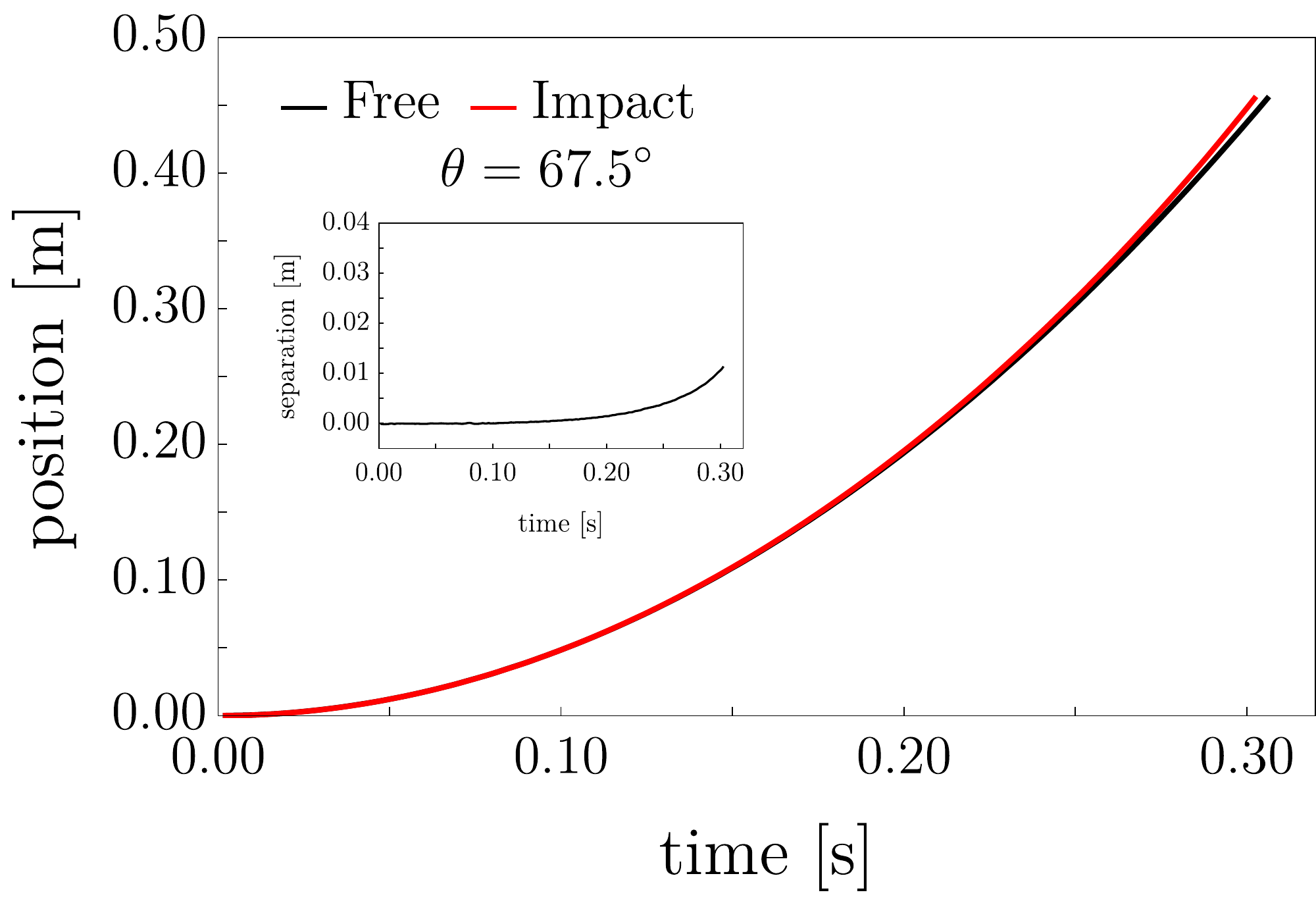}
	\end{minipage}
	\begin{minipage}[t]{17cm}
		\includegraphics[width=0.48\linewidth,valign=b]{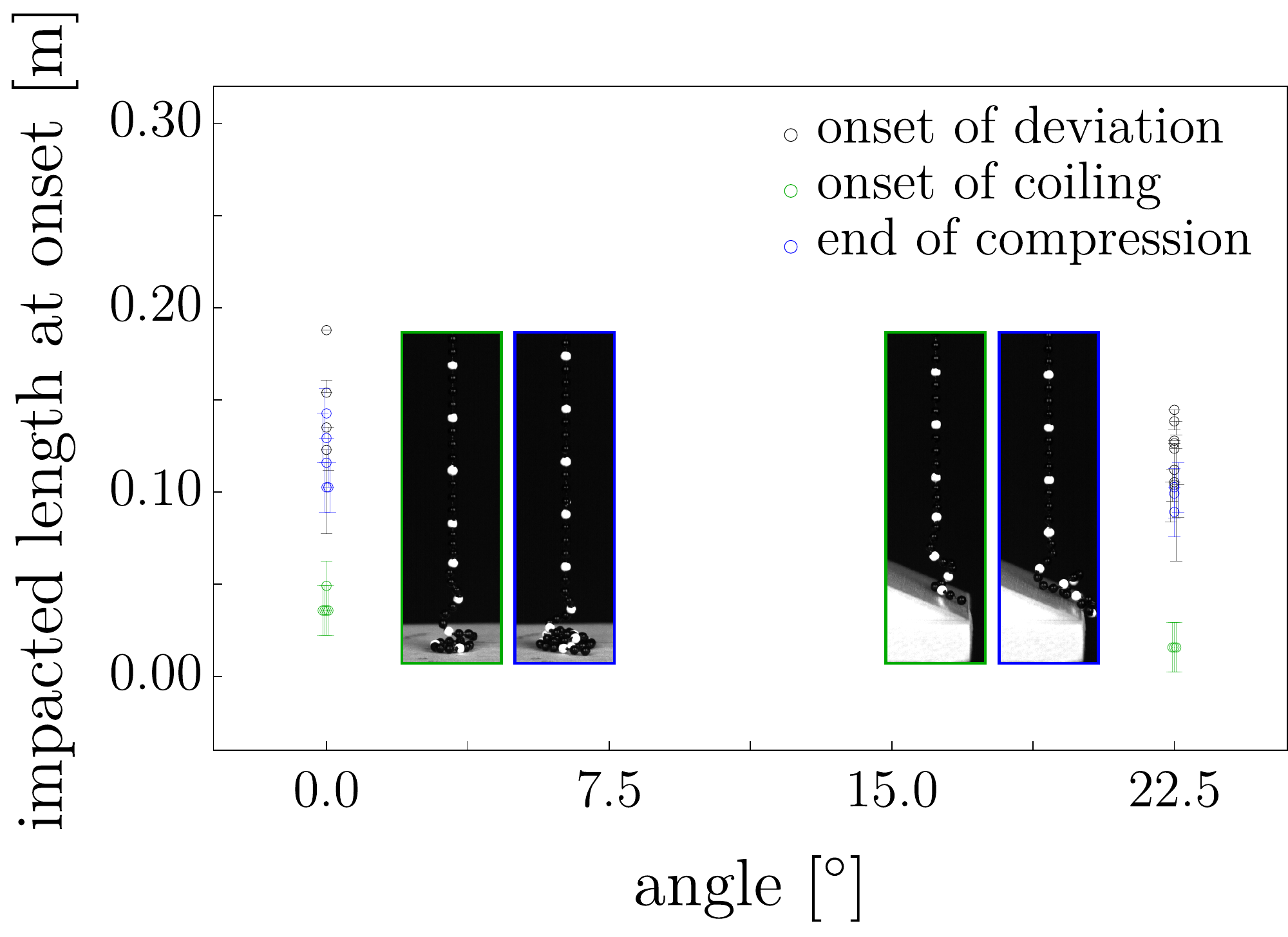}
		\hfill
		\includegraphics[width=0.48\linewidth,valign=b]{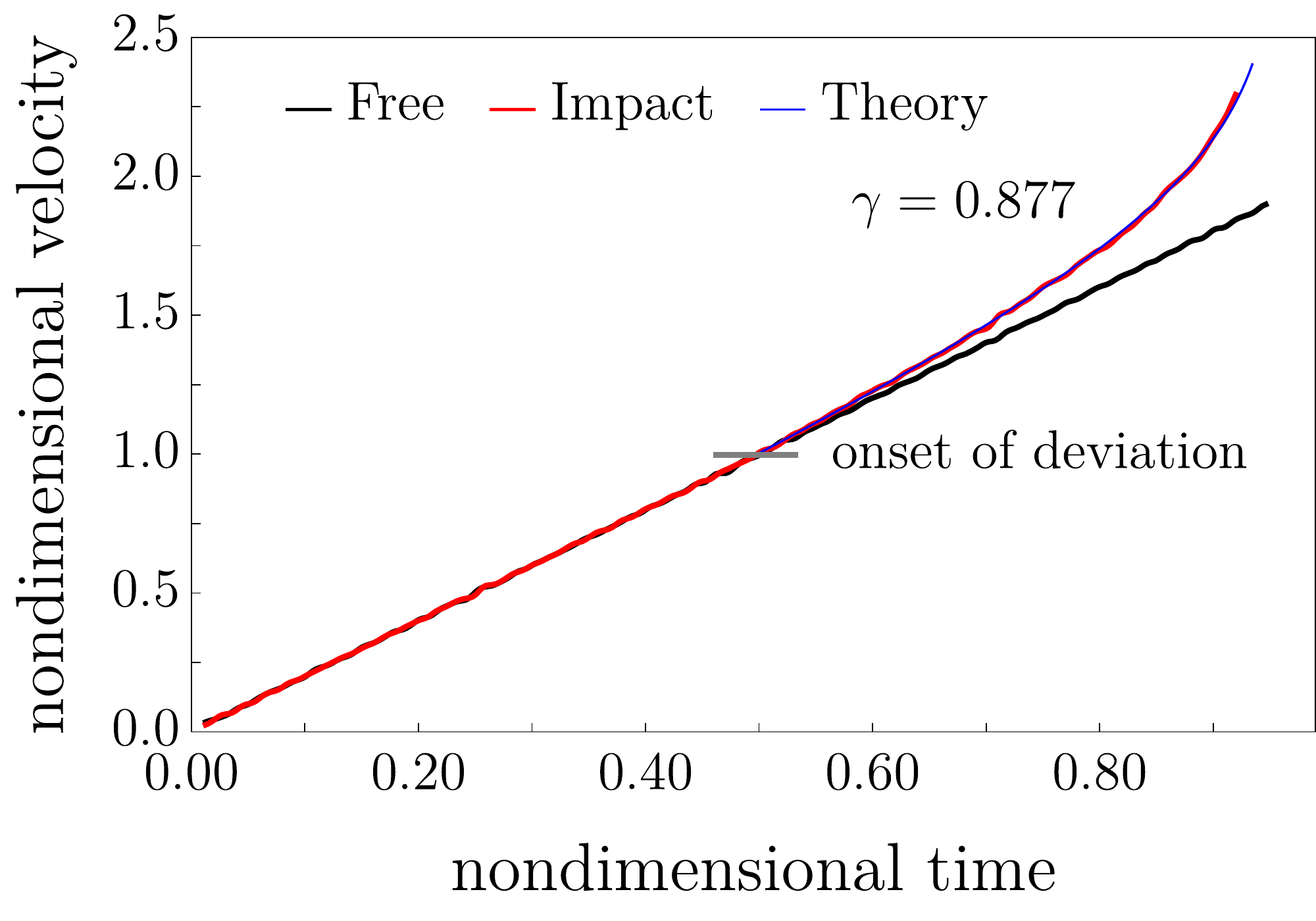}
	\end{minipage}
	\caption{Top left and right: As in Figure \ref{Fig:Experimental_setup}, positions of markers near the top ends of freely falling (black) and impeded (red) chains, and their differences (insets), for $22.5^\circ$ and $67.5^\circ$ surfaces. At high angles, no delay in onset is observed.  Bottom left: Length of chain that has already impacted when deviation from free fall begins, when coiling begins, and when compression ends, with color-coded inset images showing the latter two events, for chains dropped onto a flat surface and a $22.5^\circ$ surface. Error bars for onset of coiling and end of compression are given by $\pm\,$0.13\,mm, one inter-white bead distance. Some data points and error bars are given slight lateral shifts to avoid overlap.  Bottom right: Velocity of freely falling (black) and impeded (red) chains for a flat surface, derived from the position data in Figure \ref{Fig:Experimental_setup}, and a theoretical solution (blue) of equation \eqref{initial_value_problem} with the constitutive parameter $\gamma$\,=\,0.877.
}
	\label{Fig:Angle}
\end{figure*}

The disappearance of the extra acceleration at intermediate angles is quite intriguing.  We find that this, along with the delayed onset phenomenon, can be explained by inspection of the micromechanical behavior near the impact point.  
From a comparison of data at low and high angles (Figure \ref{Fig:Angle}), it also appears that the delay might not exist at higher angles.
The necessary detail cannot be seen in the wide-field videos, but close-up videos of the impacts lead to several interesting observations.

We observe several stages of impact on flat and slightly inclined surfaces (\texttt{0degcloseup.mp4} and \texttt{22p5degcloseup.mp4} \cite{videos}).  An initial compression and buckling of the chain is followed by the onset of coiling.  
  However, at this stage, one can clearly see that incoming beads continue to compress as they approach the impact point.  Shortly after this, the compression process stops, and incoming links simply deflect or buckle without compressing.
As shown in Figure \ref{Fig:Angle}, it is this latter transition, rather than the onset of coiling, that correlates with the onset of extra acceleration.
The inclination of the surface seems to affect the onset of coiling and end of compression at low angles. 
At high angles, such as $67.5^\circ$ (\texttt{67p5degcloseup.mp4} \cite{videos}), the beads deflect off of the impact surface and no compression of the chain is observed at any point in the fall.  The separation between the free and impacting chains appears to exist from the beginning of impact.
At the intermediate angle of $52.5^\circ$ (\texttt{52p5degcloseup.mp4} \cite{videos}), we see that the chain compresses throughout its fall instead of deflecting, and no separation is observed.

\section{Comparison with theory}

Existing theory corresponds to impact on flat surfaces.  The arguments presented by Hamm and G\'{e}minard \cite{HammGeminard10} for a coiling chain, Ruina and co-workers \cite{Grewal11} for a link, and Virga \cite{Virga15} for a continuum containing a dissipative shock all give rise to an equation of motion for the free length $y(t)$ of the chain of the form
\begin{align}
\left(\ddot{y}+g\right)y + (1-\gamma)\dot{y}^2 = 0\, ,\label{gov_equation}
\end{align}
where $0.5 \le \gamma \le 1$ is a dimensionless constitutive parameter and $g$ the gravitational acceleration.  The primary differences between these authors' treatments is in the interpretation of $\gamma$ either as characterizing the coiling radius \cite{HammGeminard10}, inter-link contact forces \cite{Grewal11}, or the extent of internal energy dissipation \cite{Virga15}.  In prior treatments, it was assumed that the dynamics \eqref{gov_equation} held from the beginning of free-fall, but we know from our experiments that this is incorrect for impact on a flat or mildly tilted surface, so we should attempt to match only the data that is obtained after the onset of deviation.
Nondimensionalizing equation \eqref{gov_equation} using $\eta=\tfrac{y}{L}$, $\tau = t\sqrt{\tfrac{g}{2L}}$, and the derivative $()' = \tfrac{d}{d\tau}()$, we have the initial value problem
\begin{align}
\left(\eta''+2\right)\eta + (1-\gamma )\left(\eta'\right)^2 = 0 \, ,\label{initial_value_problem}
\end{align}
with initial conditions $ \eta(0) \equiv \eta_0 = \eta_{\text{exp}}$ and $\eta'(0) \equiv v_0 = \eta'_{\text{exp}}$ given by the experimentally observed position (up to an error threshold) and the predicted or experimentally observed velocity at the onset of deviation.  Fitting our data in this manner will provide different values of the constitutive parameter $\gamma$ than would be obtained by attempting, like prior authors, to fit data from the initial impact time.
 The value of $\gamma$ is sensitive to the determination of onset. 
Figure \ref{Fig:Angle} shows an example of the theory applied to the velocity curves corresponding to the position curves in Figure \ref{Fig:Experimental_setup}, with a least squares fit to $\eta'$ providing $\gamma = 0.877$ for this particular run.
Using 40 trials of flat surface data \cite{data}, for various chain lengths and drop heights, we find $\gamma$\,=\,0.88$\,\pm\,$0.01.  

The functional form \eqref{initial_value_problem} and the use of a single constitutive parameter to describe direct ball chain coiling impact, and by extension pickup \cite{Virga14, Virga15, Biggins14}, seem well supported by the data.  We note that equation \eqref{initial_value_problem} is actually integrable \cite{Virga15, PolyaninZaitsev03}.  
Substituting $F(\eta) = (\eta')^2$ in \eqref{initial_value_problem}, we obtain a first order linear 
differential equation,
\begin{align}
\left(\frac{1}{2}\frac{dF}{d\eta} +2\right)\eta + (1-\gamma)F = 0\, ,\label{linear_ODE}
\end{align} 
which may be integrated to obtain, after resubstitution with $\eta'$,
\begin{align}
\eta^{2(1-\gamma)}\left[(\eta')^2 + \tfrac{4}{3-2\gamma}\eta\right] = \left[v_0^2 + \tfrac{4}{3-2\gamma}\eta_0\right]\eta_0^{2(1-\gamma)}
\, .\label{first_integral}
\end{align} 
For a given constitutive parameter $\gamma$, trajectories specified by different initial conditions $\eta_0$ and $v_0$ depend only on the constant on the right hand side of \eqref{first_integral}.  The complicated dependencies of the constant likely arise because the original equation depends explicitly on the free length $\eta$.  A scaling may be chosen such that one or the other initial condition is normalized to unity, but not both.

\section{Concluding remarks}

The ball chain is an interesting structure in that it can behave in qualitatively different ways depending on the specifics of impact, displaying either disconnected loop or rigid link type response.
  The angles at which we observe crossovers in behavior, such as rotation rather than compression, depend on the particular chain element geometry, in this case a pair of beads linked by an internal dumbbell element.  But the fact that the transition happens at all has allowed us to reveal general principles using only this single chain type, which is fortunate as there is a surprising lack of commercially available small chains with the right balance of flexibility and mechanical connectedness to allow us to study the relevant effects. 
Enhanced accelerations from coiling or deflection are only present if the discrete rigid elements are mechanically connected, in keeping with the principle outlined by Ruina and co-workers \cite{Grewal11} and Biggins \cite{Biggins14}.
 Compression in a ball chain is akin to the breaking of contacts between loops in a link chain.  However, as proposed by Hamm and G{\'{e}}minard, the finite coiling radius of a ball chain can indeed provide the necessary torque on the falling portion of the body to induce additional acceleration.  Coiling is in fact seemingly more effective at this than the simple falling link  
  effect observed at higher angles, which would presumably correspond to a different $\gamma$ value in a model that appropriately takes into account angle of incidence.
Most importantly, this finding suggests that similar effects may be present in liquid or solid rope coiling, and so may have consequences for the stability of manufacturing processes for nonwovens or other materials \cite{PetrieDenn76, Renardy06, CruickshankMunson81, Tchavdarov93}, and for cable-laying or other partial contact scenarios involving elastic rods \cite{Zajac57, Habibi07, Habibi11, MahadevanKeller96, Jawed15}.
Finally, the data support the use of a continuum description of this discrete system of the form \eqref{initial_value_problem}, which in the context of Virga's derivation corresponds to a prescription of dissipation at the impact discontinuity that scales quadratically with velocities \cite{Virga15}.
However, while the determination of $\gamma$\,=\,0.88$\,\pm\,$0.01 seems quite robust for this chain type undergoing direct impacts, this parameter alone does not encode enough information to describe general oblique impact, particularly as more than one mechanism can be operative.
We expect that chains can be treated as simple string-like continua with shocks, at which obtain complicated constitutive relations between singular sources of linear and angular momentum and energy, whose dependencies on the body and impact geometry can either be obtained experimentally or derived from micromechanical arguments \cite{OReillyVaradi99, SinghHanna17}.

Why does the end of compression differ from the onset of coiling, and why does it occur after a characteristic length of chain has undergone impact?  We don't know, but suspect that this also depends on the particular chain type.

\section*{Acknowledgments}
We thank K E Daniels, J-C G{\'{e}}minard, E Hamm, J Papadopoulos, and A Ruina for helpful suggestions and discussions, D Aliaj for early work on the experimental setup, A Patalano for help with PTVlab, and H G Wood for donation of a sacrificial rectangle. We acknowledge support from U.S. National Science Foundation grant CMMI-1462501.

\bibliographystyle{unsrt}

\end{document}